\documentclass[12pt, letterpaper]{article} 

\usepackage[top=25mm, bottom=25mm, left=25mm, right=25mm]{geometry}

\usepackage{setspace}
\usepackage{graphicx}

\usepackage{natbib} \bibpunct{(}{)}{;}{author-year}{}{,}

\begin{document} 

\begin{center} 
{\Large\bf SLiM: Simulating Evolution with Selection and Linkage}
\vskip 1cm 
{\bf Philipp W. Messer} 
\vskip 1cm 
{\it Department of Biology, Stanford University, Stanford, CA 94305 } 
\end{center}

\vskip 1cm

\doublespacing

\begin{flushleft}

{\bf Running title:}\\
Simulating linked selection
\vskip 1cm


\vskip 1cm

{\bf Corresponding author:}\\
\doublespacing
Philipp W. Messer\\ 
Department of Biology\\ 
Stanford University\\ 
371 Serra Mall\\
Stanford, CA, 94305\\ 
phone: +1 650 736 4952\\ 
fax: +1 650 723 6132\\ 
email: messer@stanford.edu 

\end{flushleft}

\newpage

\section*{ABSTRACT}

SLiM is an efficient forward population genetic simulation designed for studying the effects of linkage and selection on a chromosome-wide scale. The program can incorporate complex scenarios of demography and population substructure, various models for selection and dominance of new mutations, arbitrary gene and chromosomal structure, and user-defined recombination maps.

\vspace{2cm}

\noindent{\bf Availability:} SLiM is an easy-to-use C++ command line program freely available under the GNU GPL license from http://www.stanford.edu/$\sim$messer/software. A comprehensive documentation for SLiM can be downloaded from the program website.

\newpage

In a forward simulation every individual in the population is followed explicitly. Although this is computationally more intensive than coalescent simulations~\citep{Hudson2002}, it remains prerequisite for modeling the complex effects of linked selection, such as background selection, Hill-Robertson interference, and genetic draft. Recent findings suggest that linked selection can profoundly alter the patterns of genetic variation in many eukaryotic species~\citep{Sella2009,Lohmueller2011,Weissman2012,Messer2012a}. Importantly, the extent of these effects appears to vary strongly between different chromosomal regions due to local differences in recombination rate, functional density, as well as the rate and strength of positive selection. Efficient forward simulations capable of incorporating linkage and selection in scenarios with realistic gene and chromosome structure are hence crucial for investigating the potential consequences of linked selection on population genetic theory and methods~\citep{Messer2012a}. 

Forward simulations have a long-standing tradition in population genetics and various programs have been developed~\citep{Carvajal2010,Hoban2011}. For any such program, there is a trade-off between efficiency and flexibility. Simulations based on combined forward-backward approaches, such as MSMS~\citep{Ewing2010} and forwsim~\citep{Padhukasahasram2008}, can be very fast yet at the cost that they remain limited to scenarios with only a single selected locus. The available programs that allow to model scenarios with multiple linked selected polymorphisms, such as FREGENE~\citep{Chadeau2008},  GENOMEPOP~\citep{Carvajal2008a}, simuPOP~\citep{Peng2005}, or SFS\_CODE~\citep{Hernandez2008}, either lack the ability to model realistic gene and chromosome structure or are not efficient enough to allow for simulations on the scale of entire eukaryotic chromosomes.

Here we present SLiM, a population genetic simulation targeted at bridging the gap between efficiency and flexibility for studying linked selection in models with realistic gene and chromosome structure. Special emphasis was further placed on the ability to model and track individual selective sweeps -- both complete and partial. While retaining all capabilities of a forward simulation, SLiM utilizes sophisticated state of the art algorithms and optimized data structures to achieve high computational performance. The program enables simulations on the scale of entire eukaryotic chromosomes in reasonably large populations, allowing to study linkage effects on polymorphism and divergence in unprecedented detail. The program website provides a comprehensive documentation of SLiM that includes several example applications, such as modeling the hitchhiking of deleterious mutations under recurrent selective sweeps, the effects of background selection with realistic gene structure, and adaptive introgression after a population split.

SLiM simulates the evolution of diploid genomes in a population of hermaphrodites under a Wright-Fisher model with selection (Figure 1A). In each generation, a new set of offspring is created, descending from the individuals in the previous generation. The probability of becoming a parent is proportional to an individual's fitness, which is determined by the selection and dominance effects of the mutations present in its genomes. Gametes are generated by recombining parental chromosomes and adding new mutations. 

Mutations can be of different user-defined mutation types, specified by the distribution of fitness effects (DFE) and dominance coefficient; examples could be synonymous, adaptive, and lethal mutations. The possibility to specify arbitrary dominance effects allows for modelling a variety of evolutionary scenarios, including balancing selection and recessive deleterious mutations. Genomic regions can be of different user-defined genomic element types, specified by the particular mutation types that can occur in such elements and their relative proportions; examples could be exon and intron.

While every mutation has a specific position along the chromosome, the simulation makes an infinite sites assumption in the sense that a chromosome can harbor more than one mutation at the same site and that back-mutations do not occur. SLiM does not model the actual nucleotide states of mutations and assumes a shift model of selection implying that, once a particular mutation has become fixed in the population, the fitness at this site is reset to one. Fixed mutations are then removed from the population and recorded as substitutions.

SLiM can model complex scenarios of demography and population substructure. The simulation allows for arbitrary numbers of subpopulations to be added at user-defined times, either initialized with new individuals, or with individuals drawn from another subpopulation when modeling a population split or colonization event. Subpopulation sizes can be changed at any time, allowing for demographic events such as population bottlenecks or expansions. Migration rates can further be specified and changed for any pair of subpopulations independently. 
     
In order to establish genetic diversity, simulations first have to undergo a burn-in period. Alternatively, simulations can be initialized from a set of pre-defined genomes provided by the user. This could be, for instance, the output from a previous simulation run. The user can further specify predetermined mutations to be introduced at specific time points. Such mutations can be used to investigate individual selective sweeps or to track the frequency trajectories of particular mutations in the population. Predetermined adaptive mutations can also undergo only partial selective sweeps, where positive selection ceases once the mutation has reached a particular population frequency smaller than one.  

SLiM provides a variety of output options including: (i) the complete state of the population at specified time points -- in terms of all mutations and genomes present in the population; (ii) random samples of specific size drawn from a subpopulation at given time points; (iii) lists of all mutations that have become fixed, together with the times when each mutation became fixed; (iv) frequency trajectories of particular mutations over time. 

In order to benchmark SLiM, we compared its runtime with that of SFS\_CODE~\citep{Hernandez2008}, a popular forward simulation of similar scope. We simulated a chromosome of length $L$ with uniform mutation rate $u$ and recombination rate $r$ in a population of size $N$ over the course of $10N$ generations, assuming an exponential DFE with $2N\overline{s} = -5$. For the base scenario, we chose $L=5$ Mbp, $N=500$, $u=10^{-9}$, and $r=10^{-8}$ per site per generation. We then varied these four parameters independently to analyze how each individually affects the runtimes of the two programs. Since SFS\_CODE is limited to simulating loci smaller than 10 Mb, we simulated a corresponding number of linked loci of length 1 Mb each. Simulations were conducted on a standard iMac desktop with a 2.8 Ghz Intel core 2 Duo CPU and 4 GB of memory. Figure 1B shows the measured runtimes of the two programs. In all scenarios, SLiM outcompetes SFS\_CODE by a substantial margin, typically requiring five to ten times shorter simulation times.   

The computational efficiency of SLiM enables simulations on the scale of entire eukaryotic chromosomes in reasonably large populations. For example, simulating the functional regions in a chromosome of length $L=10$ Mb over $10^5$ generations in a population of size $N=10^4$ with $u=10^{-8}$ and $r=1$ cM/Mb, assuming a functional density of 5\% and an exponential DFE with $2N\overline{s}=-10$, takes approximately 10 hours on a single core. Note that the values for $N$, $u$, and $r$ in this simulation resemble commonly used estimates for human evolution.

\section*{Acknowledgements}

The author would like to thank Dmitri Petrov for the initial motivation to devise this program, as well as Zoe Assaf, David Enard, Peter Keightley, and Nandita Garud for testing the program and helpful comments. Part of this research was funded by the National Institutes of Health (grants GM089926 and HG002568 to Dmitri Petrov).

\newpage

\bibliographystyle{genetics}
\renewcommand\refname{Literature Cited} 
{\bibliography{../../bibliography/bibliography}}

\newpage

\begin{center}
\begin{figure}[h!]
\includegraphics[width=\textwidth]{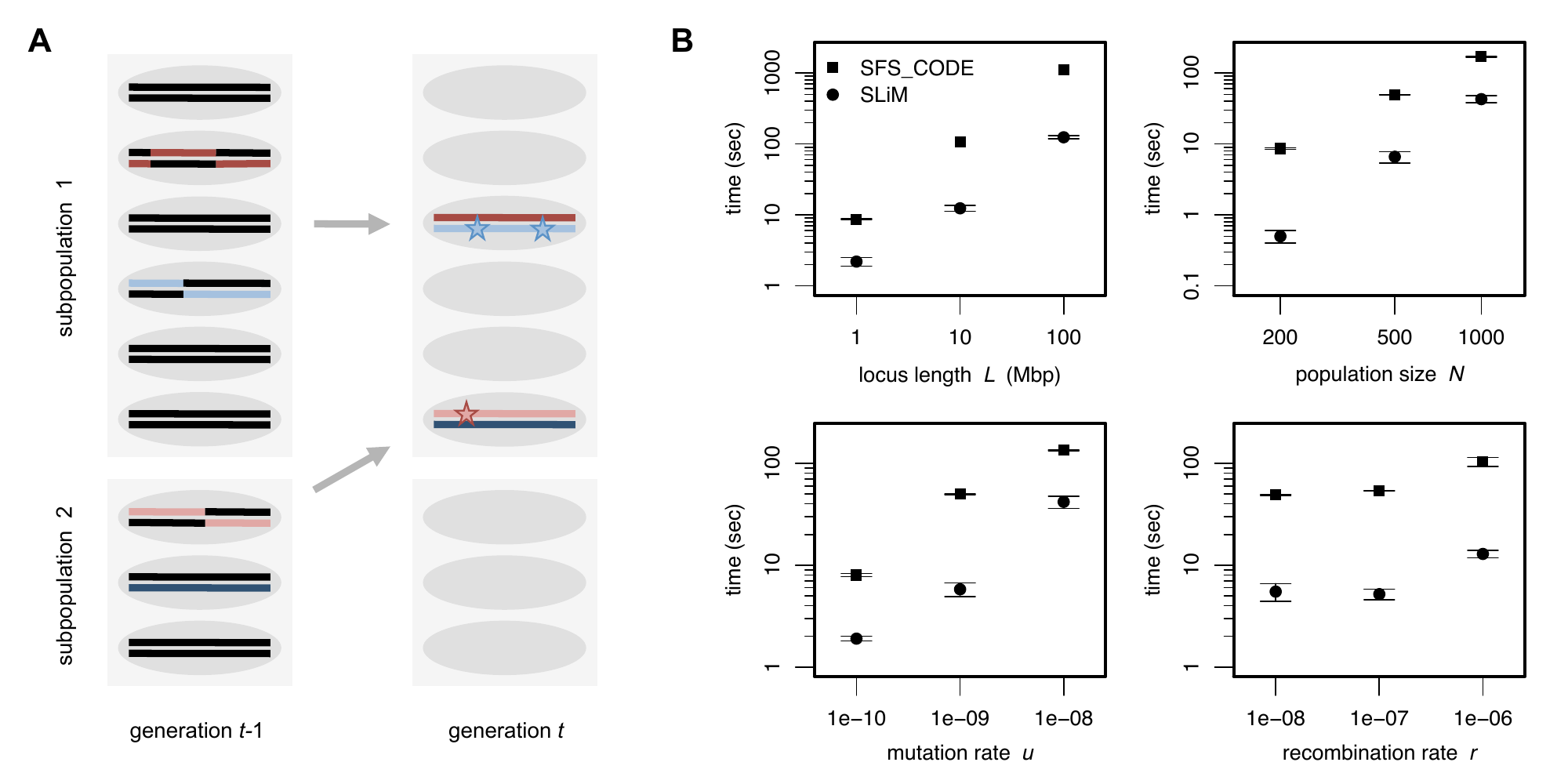}
\end{figure}
\end{center}

\noindent {\bf Figure 1} (A) Illustration of SLiM's core algorithm for a scenario with two subpopulations, the first consisting of six individuals and the second consisting of three individuals. In each generation, a new set of offspring is created, descending from the individuals in the previous generation. The parent individuals are drawn with probabilities being proportional to their fitness. Gametes are then generated by recombining parental chromosomes and adding new mutations according to the specified mutational parameters. For migrating individuals, the parents are chosen from the respective source subpopulation. Once all offspring are created, they become the parents for the next generation. (B) Comparison of the observed runtimes of SLiM and SFS\_CODE under different evolutionary scenarios. In each panel, the parameter on the x-axis was varied while the others were kept constant at their respective base values ($L=5$ Mbp, $N=500$, $u=10^{-9}$, and $r=10^{-8}$). Data points show the means, error bars the standard deviations measured over 10 simulation runs.

\end{document}